\documentclass[a4paper,11pt]{article}
\pdfoutput=1
\usepackage{jheppub}
\usepackage{braket}

\title{Thermodynamic phase transitions from dynamical compact dimensions}
\author{Zach Polonsky$^{*\dag}$} \emailAdd{polonsza@mail.uc.edu}
\author{and Alex Flournoy$^*$} \emailAdd{aflourno@mines.edu}
\affiliation{$^*$Department of Physics, Colorado School of Mines, 1500 Illinois St, Golden, CO 80401, United States \\
$^\dag$Department of Physics, University of Cincinnati, 2600 Clifton Ave, Cincinnati, OH 45221, United States
}
\date{}

\abstract{
We perform a double quotient of global AdS$_4$ and study its thermal properties. We find that a double quotient yields a spacetime with an expanding compact dimension. Studying the entanglement of the dual CFT we find that, at early times, the spacetime has thermal properties which disappear after a critical time.
For slow expansion, this critical time depends on the expansion rate as expected, but becomes much more sensitive with more rapid expansion rates.}

\begin{document}
\maketitle
\flushbottom
\section{Introduction}

Over two decades ago, a new solution to Einstein's equations in $2+1$ dimensions with a negative cosmological constant was discovered by Ba\~{n}ados \textit{et al} \cite{btz}. This solution, known as the BTZ black hole, is an eternal black hole spacetime which is locally identical to Anti-de Sitter (AdS) space. It was also shown that this spacetime could be obtained from a quotient of global AdS \cite{btz,btzlong}.

The BTZ black hole has proven to be an invaluable tool for studying quantum gravity, especially in regards to the AdS/CFT correspondence due its exhibiting interesting global properties, while remaining locally trivial \cite{eternalbh,ryutak,hrt}. While the BTZ black hole has been generalized to higher dimensions, calculations in these higher-dimensional generalizations are made more complicated due to time-dependences in the spacetimes.

Of interest to this paper is the case of a double quotient of global AdS in $3+1$ dimensions. In $2+1$ dimensions, it has been shown that, depending on the identification chosen, the second quotient either produces a third exterior region and creates a multi-boundary wormhole, or identifies the two asymptotic boundaries of the BTZ black hole to create a single-exterior black hole \cite{brill,loukomarolf,louko}. The implications of this in terms of multi-partite entanglement of $(1+1)$-dimensional CFTs has also been discussed \cite{multibound}. Where a single quotient of $(3+1)$-dimensional global AdS yields an exotic single-exterior black hole, we find that a double quotient results in a four-exterior black hole spacetime reminiscent of a big bang. This is not a typical big bang, however, in the sense that the entire universe spawns from a single point. Instead, we see that only a single, compact dimension is created which then expands to become macroscopic.

We begin by reviewing the single quotient of global AdS in $3+1$ dimensions. We go on to study the thermodynamics of this spacetime using holographic methods. Then we perform the second quotient which yields a new spacetime whose thermodynamic behavior we study. We assign a Hartle-Hawking state to show that the dual CFTs are initially in an entangled thermofield double state. We then apply Ryu and Takayanagi's Holographic Entanglement Entropy (HEE) to the double-quotient spacetime to find the time-dependent behavior of the entanglement and discuss its implications on the thermodynamics of the spacetime. We end with our conclusions about the work.

\section{Single Quotient of Global AdS$_4$}

\begin{figure}
 \centering
 \includegraphics[width=.5\textwidth]{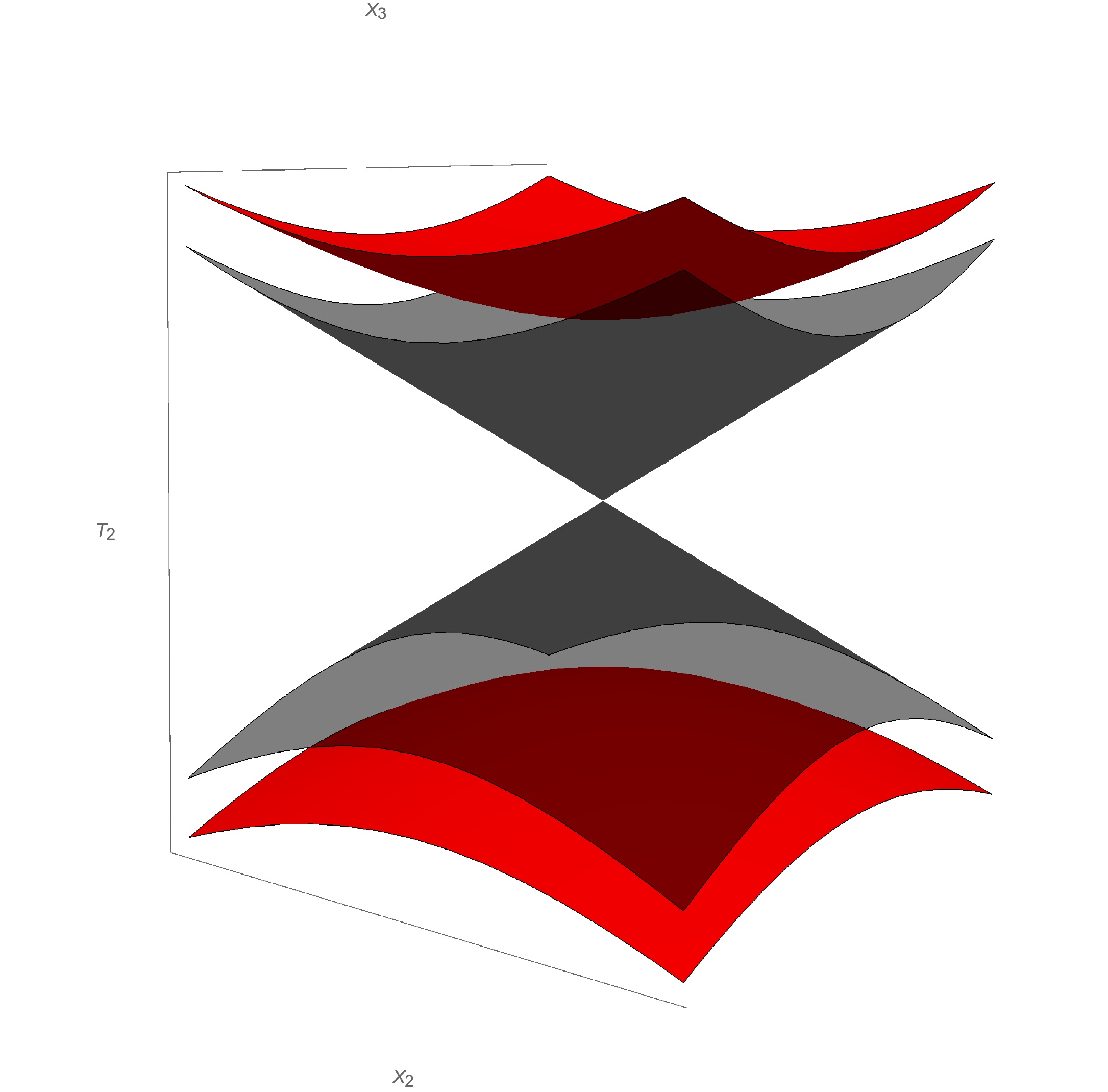}
 \caption{Singularity (red) and event horizon (black) created from a quotient by $\xi$. Here, the $T_1$ and $X_1$ coordinates are suppressed.}
 \label{blackhole}
\end{figure}

This section serves as an overview of the process used to obtain the (3+1)-dimensional topological black hole in \cite{topbh} as well introduce different coordinate systems which will be useful later on. To perform the first quotient, we begin with global $(3+1)$-dimensional AdS (AdS$_4$), defined as the surface
\begin{equation}
\label{ads4}
 -T_1^2-T_2^2+X_1^2+X_2^2+X_3^2=-l^2
\end{equation}
embedded in $\mathbb{R}^{2,3}$ with metric
\begin{equation}
\label{r23}
 ds^2=-dT_1^2-dT_2^2+dX_1^2+dX_2^2+dX_3^2
\end{equation}
where $l$ is the AdS radius. We will consider the boost-like isometry given by
\begin{equation}
\label{quovec}
 \xi=-X_1\partial_{T_1}-T_1\partial_{X_1} \quad , \quad \xi^2=-X_1^2+T_1^2
\end{equation}
to generate the quotient. To avoid closed timelike curves, we remove regions of the spacetime where $\xi^2<0$ after the quotient \cite{quo,brill}. This creates a singularity in the causal structure where timelike geodesics can end at
\begin{equation}
 -X_1^2+T_1^2=0
\end{equation}
or equivalently, by using (\ref{ads4}),
\begin{equation}
 -T_2^2+X_2^2+X_3^2=-l^2.
\end{equation}
This singularity asymptotes to the null cone given by
\begin{equation}
 T_2^2=X_2^2+X_3^2 \quad \Rightarrow \quad X_1^2-T_1^2=-l^2
\end{equation}
which we identify as the event horizon of the black hole. These surfaces are plotted in Figure \ref{blackhole}.

We can define coordinates on the surface (\ref{ads4}), which are related to the coordinates of the embedding space by
\begin{align}
 T_1=l\,\frac{1-t^2+y_1^2+y_2^2}{1+t^2-y_1^2-y_2^2}\cosh\left(\frac{r_+}{l} \phi\right) \quad &, \quad 
 T_2=\frac{2lt}{1+t^2-y_1^2-y_2^2} \nonumber \\
 \nonumber \\
 X_1=l\,\frac{1-t^2+y_1^2+y_2^2}{1+t^2-y_1^2-y_2^2}\sinh \left(\frac{r_+}{l} \phi\right) \; , \;
 X_2=&\frac{2ly_1}{1+t^2-y_1^2-y_2^2} \; , \;
 X_3=\frac{2ly_2}{1+t^2-y_1^2-y_2^2}
\end{align}
where $r_+$ is a constant. In terms of these coordinates the induced metric on the surface becomes
\begin{align}
\label{carkrusk}
 ds^2=\frac{4l^2}{(1+t^2-y_1^2-y_2^2)^2}&(-dt^2+dy_1^2+dy_2^2)
 +r_+^2\left(\frac{1-t^2+y_1^2+y_2^2}{1+t^2-y_1^2-y_2^2}\right)^2d\phi^2
\end{align}
and the Killing vector which generates the quotient becomes $\xi=\partial_\phi$ \cite{topbh}. The quotient makes the identification $\phi=\phi+2\pi$, and the coordinate ranges after the quotient are given by $t,y_i\in(-\infty,\infty)$ and $\phi\in[0,2\pi)$. In these coordinates the singularity is given by the surface
\begin{align}
 &\xi^2=l^2 \,\left(\frac{1-t^2+y_1^2+y_2^2}{1+t^2-y_1^2-y_2^2}\right)^2=0 \quad \rightarrow \quad -t^2+y_1^2+y_2^2=-1
\end{align}
and the event horizon is given by
\begin{align}
 &T_1^2-X_1^2=l^2\,\left(\frac{1-t^2+y_1^2+y_2^2}{1+t^2-y_1^2-y_2^2}\right)^2=l^2 \quad \rightarrow \quad -t^2+y_1^2+y_2^2=0.
\end{align}
We can also see that the embedding coordinates diverge at the surface
\begin{equation}
 1+t^2-y_1^2-y_2^2=0
\end{equation}
which we associate with the boundary of the spacetime. Here we note that, aside from the singularity, the spacetime in these coordinates is geodesically complete, and therefore no further maximal extension is necessary \cite{topbh}. In addition we see that this spacetime has only a single, connected asymptotic boundary, as opposed to the $(2+1)$-dimensional case, which has two asymptotically distinct boundaries \cite{btz,topbh,btzlong}. 

For future reference, it is also helpful to define ``Schwarzschild coordinates'' which cover the full exterior of the black hole. These can be defined by

\begin{align}
\label{fullexttrans}
 t=\chi \sinh\, t' \quad ,\quad y_1 = \chi \cos \psi\, \cosh\, t' \quad , \quad y_1 = \chi \sin \psi\, \cosh\, t'
\end{align}
giving the metric \cite{topbh}

\begin{equation}
    ds^2=\frac{4l^2}{\left(1-\chi^2\right)^2}\left(-\chi^2dt'^2+d\chi^2+\chi^2\cosh^2t'\,d\psi^2\right)+r_+^2\left(\frac{1+\chi^2}{1-\chi^2}\right)^2d\phi^2.
    \label{fullext}
\end{equation}

The first quotient of AdS$_4$ is an interesting case to study due to its single asymptotic boundary. Once we perform the second quotient of the spacetime, this single boundary will be split into four distinct asymptotic boundaries, and will also exhibit non-trivial thermal behavior. 

\section{Particle Modes in the Boundary Theory}
\label{firstquosec}

It has been shown \cite{maldacena1,eternalbh,ryutak,hrt} that the thermodynamics exhibited by asymptotically AdS spacetimes are mirrored in CFTs defined on the conformal boundary of the spacetime. We will use this to our advantage to track the thermal behavior of the double-quotient spacetime in the next section. First, we will define global coordinates on AdS$_4$ in terms of the embedding coordinates by

\begin{align}
 T_1=l\frac{1+r^2}{1-r^2}&\cos t\quad , \quad 
 T_2=l\frac{1+r^2}{1-r^2}\sin t \nonumber \\
 \nonumber \\
 X_1=\frac{2lr}{1-r^2}\cos\lambda \quad , \quad 
 X_2&=\frac{2lr}{1-r^2}\cos\psi\,\sin\lambda \quad , \quad
 X_3=\frac{2lr}{1-r^2}\sin\psi\,\sin\lambda
\end{align}
with ranges $r\in[0,1)$, $\psi\in[0,2\pi)$, $\lambda\in[0,\pi]$, and $t\in(-\infty,\infty)$ after enforcing a universal covering to avoid closed timelike curves \cite{ads}. The metric becomes

\begin{align}
\label{gloads}
 ds^2=\frac{4l^2}{\left(1-r^2\right)^2}\left[-\frac{\left(1+r^2\right)^2}{4}\,dt^2 +dr^2+r^2d\lambda^2+r^2\cos^2\lambda\, d\theta^2\right]
\end{align}
where the boundary is located at $r=1$. The Killing vector that generates the quotient to form the black hole is given by (\ref{quovec}). There is another Killing vector orthogonal to $\xi$ given by $\eta=X_2\partial_{T_2}+T_2\partial_{X_2}$, where the motivation for introducing $\eta$ will become clear. To get the metric on the boundary, we multiply (\ref{gloads}) by the conformal factor

\begin{equation}
 \Omega^2=\frac{\left(1-r^2\right)^2}{4l^2}
\end{equation}
and take $r\to1$ to obtain

\begin{equation}
 d\sigma^2=-dt^2+d\lambda^2+\cos^2\lambda\,d\psi^2.
\end{equation}

The conformal Killing vectors on the boundary corresponding to $\xi$ and $\eta$ are given by

\begin{align}
 \xi_b=\cos\lambda\,\sin t\,\partial_t+\cos t\,&\sin\lambda\,\partial_\lambda \nonumber \\
 \eta_b=\cos t\,\sin\lambda\,\cos\psi\,\partial_t+&\sin t\,\cos \lambda \,\cos \psi \partial_\lambda-\sin t\,\csc \lambda \,\sin\psi \, \partial_\psi
\end{align}
respectively \cite{ckv}. To find the region of the boundary that will survive the quotient, we must find the region where $\xi_b^2>0$. This region corresponds to the single diamond

\begin{equation}
 0<\lambda<\pi, \quad |t|<\frac{\pi}{2}-\left|\lambda-\frac{\pi}{2}\right|
\end{equation}
in contrast to the BTZ black hole, which gives two separate diamonds \cite{loukomarolf,louko}. This is a reflection of the fact that the $(3+1)$-dimensional topological black hole has a single, connected boundary.

To better understand the action of $\xi_b$ on the boundary, we define the coordinates (inspired by those used in \cite{loukomarolf})

\begin{align}
 \alpha&=-\log\left[\tan\left(\frac{\lambda-t}{2}\right)\right] \nonumber \\
 \beta&=\log\left[\tan\left(\frac{\lambda+t}{2}\right)\right]
\end{align}
giving the metric

\begin{equation}
\label{notinv}
 d\sigma^2=\frac{d\alpha d\beta +\cosh^2\left[(\alpha+\beta)/2\right]\,d\psi^2}{\cosh\alpha\,\cosh\beta}
\end{equation}
where the conformal Killing vectors are now given by

\begin{align}
 \xi_b&=-\partial_\alpha+\partial_\beta \nonumber \\
 \eta_b&=\cos\psi\,(\partial_\alpha+\partial_\beta)-\sin\theta\tanh\left(\frac{\alpha+\beta}{2}\right)\partial_\psi.
\end{align}
The vector $\xi_b$ maps the point 

\begin{equation}
(\alpha,\beta,\psi)\to(\alpha-c,\beta+c,\psi)
\end{equation}
where $c$ is a constant. Unfortunately, the metric (\ref{notinv}) is not invariant under an action generated by $\xi_b$, but the conformally related metric

\begin{equation}
 d\sigma^2=-d\alpha\,d\beta+\cosh^2\left(\frac{\alpha+\beta}{2}\right)d\psi^2
\end{equation}
is. Finally, defining $\alpha=t'-r_+\phi$ and $\beta=t'+r_+\phi$ where $\tau,\phi\in(-\infty,\infty)$, we obtain the metric

\begin{equation}
\label{bound}
 d\sigma^2=-dt'^{\,2}+\cosh^2t'\,d\psi^2+r_+^2d\phi^2
\end{equation}
and the conformal Killing vectors take the form

\begin{align}
 \xi_b&=\partial_\phi \nonumber \\
 \eta_b&=\cos\psi\,\partial_{t'}-\sin\psi\,\tanh t' \, \partial_\psi.
\end{align}
A quotient by $\xi_b$ makes the identification $\phi=\phi+2\pi$ and we see that the metric (\ref{bound}) is the conformal boundary of the topological black hole in the coordinates (\ref{fullext}).

Near $t'=0$, we find $\eta_b=\cos\psi\, \partial_{t'}$ is purely timelike, except for the points $\psi=\pm\pi/2$. Moreover, for the region

\begin{equation}
 D_f\doteq\,\left\{\psi\in\left(-\frac{\pi}{2},\frac{\pi}{2}\right)\right\}
\end{equation}
$\eta_b$ is future-directed, while for the region 

\begin{equation}
 D_p\doteq\, \left\{\psi\in\left(\frac{\pi}{2},\frac{3\pi}{2}\right)\right\}
\end{equation}
$\eta_b$ is past-directed. Therefore, we naturally associate $\eta_b$ with particle modes in the boundary theory, where positive energy modes are associated with $\eta_b$ in $D_f$ and negative energy modes are associated with $\eta_b$ in $D_p$.

As the system evolves away from $t'=0$, we find that $\eta_b$ is no longer timelike over the full boundary and thus $\eta_b$ will only generate particles on the region of the spacetime corresponding to

\begin{equation}
 \cosh^2t'\,\cos^2\psi-\sinh^2t'>0.
\end{equation}
Looking at the form of (\ref{fullexttrans}), we can see that this is just the region of the boundary $y_1^2-t^2>0$.

We would now like to show that the regions $D_f$ and $D_p$ are in pure states. Since these regions have a global timelike Killing vector associated with them, we can use the Ryu-Takayanagi minimal-area holographic entanglement entropy proposal to find the entanglement entropy of these regions \cite{ryutak,aspects}.

The holographic entanglement entropy conjecture of Ryu and Takayanagi states that the entanglement entropy, $S_A$ of a spatial region $A$ of a CFT can be calculated from the area of the minimal area surface, $\gamma$, in the dual spacetime which terminates on the boundary of $A$. This relationship is given by
\begin{equation}
S_A=\frac{\mathrm{Area}(\gamma)}{4G^{(d)}_N}
\end{equation}
where $G^{(d)}_N$ is the $d$-dimensional gravitational constant \cite{ryutak,aspects}.

We can now find the entanglement entropy of the region of the boundary where $\eta_b$ is timelike by simply relating coordinates (\ref{carkrusk}) to Poincar\'{e} coordinates \cite{ryutak,hrt,aspects}. The metric of global AdS$_4$ in Poincar\'{e} coordinates is given by
\begin{equation}
 ds^2=\frac{l^2}{Z^2}\left(dW_+dW_-+dY^2+dZ^2\right)
\end{equation}
where $W_{\pm},Y\in(-\infty,\infty)$ and $Z\in(0,\infty)$ \cite{poincare}. Note that this is the null coordinate form of the Poincar\'{e} metric where $W_\pm=X\pm T$ with $T$ and $X$ being timelike and spacelike coordinates, respectively. The boundary of the spacetime is at $Z=0$. If we consider a strip on the boundary at fixed time given by
\begin{equation}
 W_+-W_-=const \; \Rightarrow \; \Delta W_+-\Delta W_-=0,
\end{equation}
the length of the strip will be given by $\Delta Y=L$ and the width will be given by
\begin{equation}
\label{rsq}
 R^2=\Delta W_+ \Delta W_-.
\end{equation}

\begin{figure}
 \centering
 \includegraphics[width=.3\textwidth]{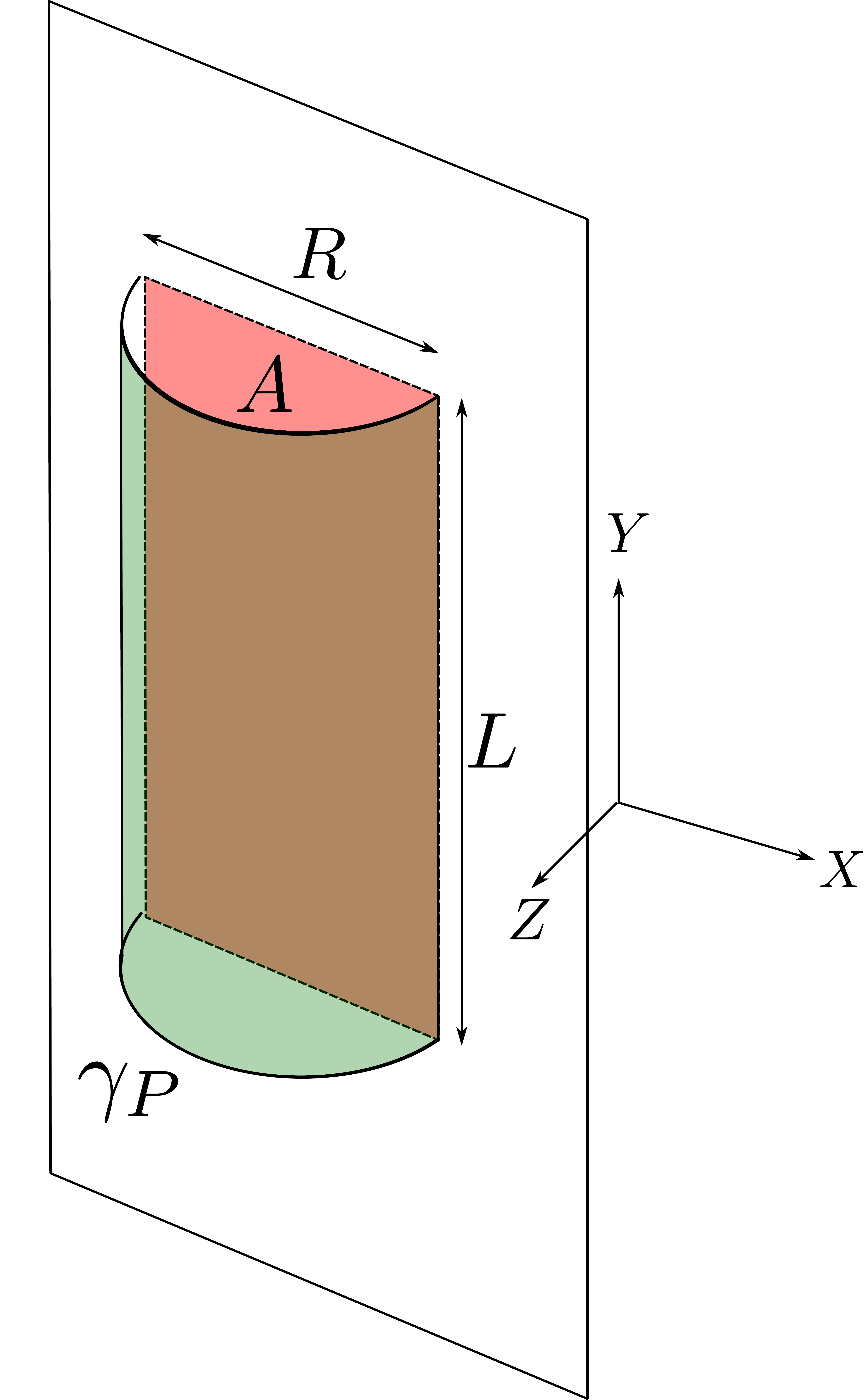}
 \caption{Minimal-area surface, $\gamma_P$, (green) corresponding to the strip, $A$, (red) at a constant-time slice of global AdS$_4$.}
 \label{strip}
\end{figure}

If we choose a minimal-area surface, $\gamma_P$, which terminates only on the boundary in the $X=1/2(W_++W_-)$ direction as shown in Figure \ref{strip}, the area of $\gamma_P$ will be given by
\begin{equation}
\label{poinarea}
 \mathrm{Area}(\gamma_P)=2l^2\left(\frac{L}{\epsilon}\right)-\kappa l^2\left(\frac{L}{R}\right)
\end{equation}
where $\epsilon$ is a cutoff introduced to prevent the expression from diverging and
\begin{equation}
\kappa=4\pi \left(\frac{\Gamma\left(\frac{3}{4}\right)}{\Gamma\left(\frac{1}{4}\right)}\right)^2
\end{equation}
is a positive constant \cite{hrt,aspects}. For the expression (\ref{poinarea}) to be consistent, we can see that the area of the surface $\gamma_P$ must vanish when the region, $A$, on the boundary also vanishes, i.e. Area($\gamma_P$)$\to 0$ as $\epsilon,R\to 0$.

Now, we can relate the Poincar\'{e} coordinates to those used in (\ref{carkrusk}), but to ease the calculations, we will introduce a polar form of the coordinates with $y_1=r\cos\psi$ and $y_2=r\sin\psi$. These coordinates are related to the null form of Poincar\'{e} coordinates by

\begin{align}
 W_{\pm}&=\frac{2}{1-t^2+r^2}\left(r \cos\psi  \pm t\right)\,e^\phi \nonumber \\
Y&= \frac{2}{1-t^2+r^2}\; r\; \sin \psi \, e^\phi \nonumber \\
Z&=\frac{1+t^2-r^2}{1-t^2+r^2}\,e^\phi.
\end{align}

In these coordinates, the boundary is the surface where $Z=0$, which corresponds to $r=\sqrt{1+t^2}$. If we anchor the minimal-area surface, $\gamma$, to the boundary at fixed time $t=t_0$ such that $\psi\in[\psi_1,\psi_2]$ and $\phi\in[\phi_1,\phi_2]$, the cutoff at $Z=\epsilon\ll1$ corresponds to

\begin{equation}
\label{kep}
 \epsilon_{1,2}=ae^{\phi_{1,2}} \quad \mathrm{where} \quad 0<a\ll1.
\end{equation}
For simplicity, we will center the region $A$ on the boundary such that $\psi_1=-\Psi$ and $\psi_2=\Psi$. Thus, the length, $L$, of $A$ is given by

\begin{equation}
\label{kl}
 L=\Delta Y=\sqrt{1+t_0^2}\left(e^{\phi_2}+e^{\phi_1}\right)\sin\Psi
\end{equation}
and the width, $R$, is given by (\ref{rsq})

\begin{equation}
\label{krsq}
 R^2=\left(e^{\phi_2}-e^{\phi_1}\right)^2\left[\left(1+t_0^2\right)\cos^2\Psi-t_0^2\right].
\end{equation}

Using (\ref{kep}), (\ref{kl}), and (\ref{krsq}) in (\ref{poinarea}), we find an expression for the area of $\gamma$ using $\epsilon^2=\epsilon_1 \epsilon_2$

\begin{align}
 \label{areakrusk}
 \mathrm{Area}(\gamma)=\,4l^2\left(\frac{\cosh \Delta}{a}\right)\sqrt{1+t_0^2}\sin \Psi-\kappa l^2\left(\frac{\sqrt{1+t_0}\sin\Psi}{\sqrt{(1+t_0^2)\cos^2\Psi-t_0^2}}\right)\coth \Delta
\end{align}
where $\Delta=\left(\phi_2-\phi_1\right)/2$. Here, we recall that the area must vanish when both $\epsilon,R\to0$, corresponding to

\begin{equation}
 a=0 \quad ,\quad \left(1+t_0^2\right)\cos^2\Psi-t_0^2=0.
\end{equation}
We notice that the latter of these requirements is equivalent to $y_1^2-t^2=0$, which is exactly the boundary of $D_f$. Since the area of $\gamma$ vanishes on the boundary of $D_f$, we can conclude that the CFT on $D_f$ is in a pure state. This also holds true for $D_p$.

The regions $D_f$ and $D_p$ will serve as useful tools for analyzing the thermodynamics of the double-quotient spacetime. This is mainly due to the fact that any CFTs defined on these regions will be ``complete'' theories i.e. we do not need any information from the rest of the boundary to describe the state of these theories. 

\section{Second Quotient of AdS$_4$}

\begin{figure}
 \centering
 \includegraphics[width=.5\textwidth]{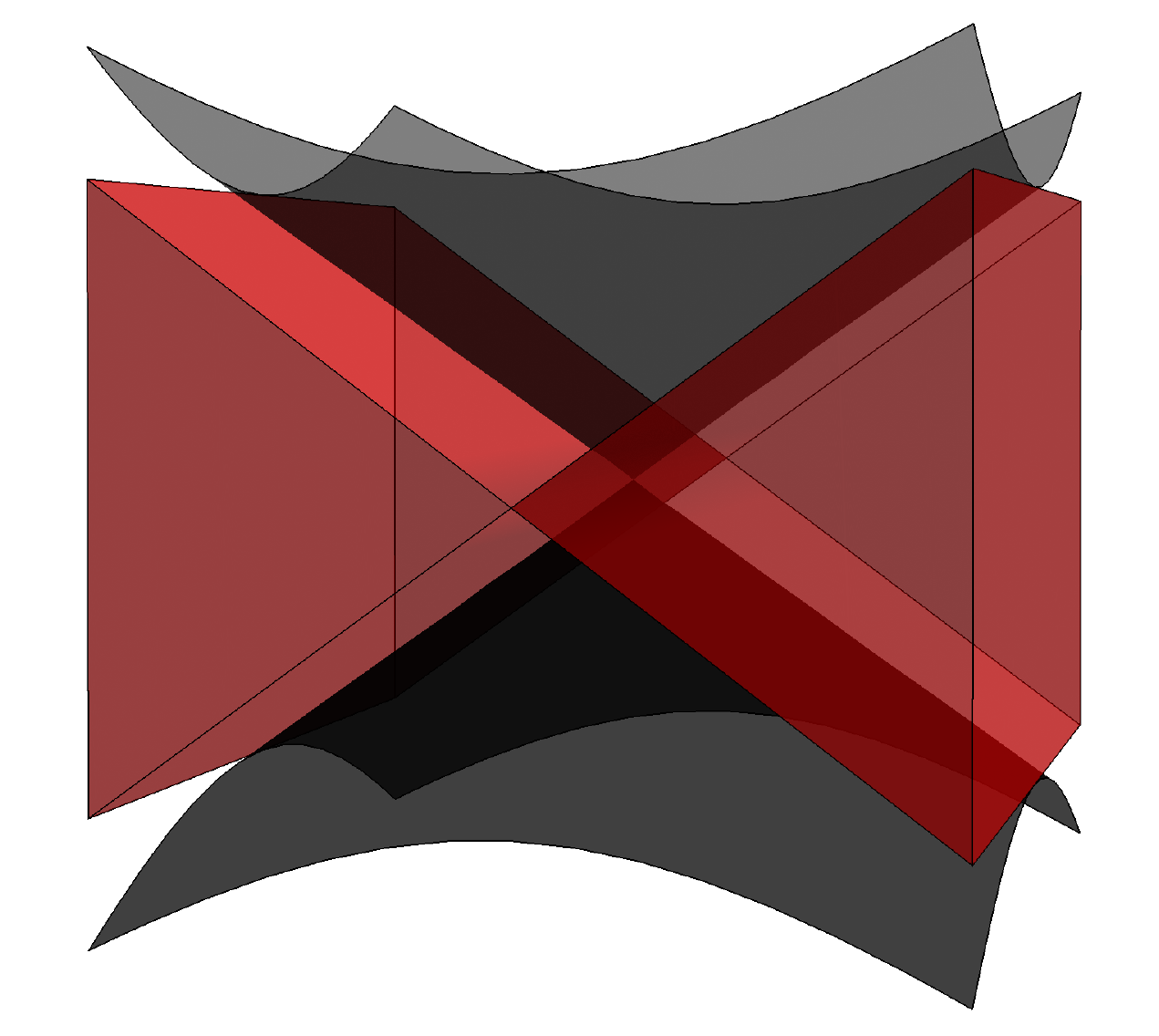}
 \caption{The regions excised during second quotient (red) end on null surfaces, running along the event horizon of the black hole (black). Since these surfaces are null, they are not bounded by event horizons and therefore, the resulting singularity is naked.}
 \label{singularity}
\end{figure}

To perform a second quotient of AdS$_4$, we first recognize that the metric (\ref{carkrusk}) has a boost-like isometry with Killing vector
\begin{equation}
\xi'=y_2 \partial_t+t \partial_{y_2}.
\label{seckvec}
\end{equation}
As we will be performing a quotient defined by this Killing vector, we will again need to excise regions of the spacetime where $\xi'^2< 0$ to avoid closed timelike curves. These regions correspond to
\begin{equation}
y_2^2-t^2>0.
\label{removed}
\end{equation}
Since the boundary of these regions are null lines $t=\pm y_2$, they are not bounded by event horizons and therefore are naked. Figure \ref{singularity} shows these naked singularities. It is also worthy to note that the introduction of these new singularities separates the spacetime into four asymptotically distinct regions: two future regions and two past regions.

We now define new coordinates
\begin{align}
t=\left|\tau\right|\cosh(\alpha\, \theta) \quad , \quad
y_2=\left|\tau\right|\sinh(\alpha\, \theta)
\end{align}
where $\theta\in(-\infty,\infty)$ and $\tau\in(-\infty,\infty)$. This transforms the metric (\ref{carkrusk}) to 
\begin{align}
ds^2=\frac{4l^2}{(1+\tau^2-y_1^2)^2}&\left(-d\tau^2+dy_1^2+\alpha^2\tau^2d\theta^2\right) +r_+^2\left(\frac{1-\tau^2+y_1^2}{1+\tau^2-y_1^2}\right)^2d\phi^2.
\label{finmet}
\end{align}
and the Killing vector (\ref{seckvec}) to
\begin{equation}
\xi'=\frac{1}{\alpha}\partial_\theta.
\label{finalkvec}
\end{equation}
This Killing vector maps the points $\theta\to\theta+a/\alpha$ where $a$ is an arbitrary parameter. Defining $\alpha=a/2\pi$, we can see that a quotient by $\xi'$ will make the $\theta$ coordinate periodic, with range $\theta\in[0,2\pi)$. The black hole singularity is now located at the surface
\begin{equation}
-\tau^2+y_1^2=-1,
\end{equation}
the black hole event horizon is given by the surface
\begin{equation}
-\tau^2+y_1^2=0,
\end{equation}
and the boundary of the spacetime is at
\begin{equation}
-\tau^2+y_1^2=1.
\end{equation}
In these coordinates, the naked singularity is located at $\xi'^2=0$ or $\tau=0$.

Since the $\theta$ coordinate is scaled by $\tau$, for very early times, the $\theta$ dimension will be microscopic and we may effectively neglect the $d\theta$ term in our metric. We are then left with the metric of the BTZ black hole \cite{eternalbh}
\begin{equation}
ds^2=\frac{4l^2}{1-y_1^2}\left(-d\tau^2+dy_1^2\right)+r_+^2\left(\frac{1+y_1^2}{1-y_1^2}\right)d\phi^2.
\end{equation}
Therefore, for times near $\tau=0$, we should expect the black hole to radiate at inverse temperature \cite{btzlong}
\begin{equation}
\beta=\frac{r_+}{2\pi}.
\end{equation}

However, away from $\tau=0$, the black hole cools to zero temperature. This can be seen by defining the coordinates
\begin{align}
\tau&=\rho \sinh(\alpha \zeta) \quad , \quad
y_2=\rho \cosh(\alpha \zeta)
\end{align}
which transform the metric (\ref{finmet}) to

\begin{align}
ds^2=\frac{4l^2}{(1-\rho^2)^2}&\left[-\alpha^2 \rho^2 d\zeta^2+d\rho^2+\alpha^2 \rho^2\sinh^2(\alpha \zeta)d\theta^2\right]
+r_+^2\left(\frac{1+\rho^2}{1-\rho^2}\right)^2d\phi^2.
\end{align}
Here, $\zeta\in (-\infty,\infty)$ and $\rho\in[0,1)$. These coordinates correspond to an observer living in one exterior of the black hole, whose event horizon is located at $\rho=0$. It is worth noting that, in these coordinates, the event horizon is 1-dimensional. We can perform a Wick rotation by taking $\zeta\to-i \zeta_E$ and the metric becomes

\begin{align}
ds^2=\frac{4l^2}{(1-\rho^2)^2}&\left[\alpha^2 \rho^2 d\zeta_E^2+d\rho^2+\alpha^2 \rho^2\sin^2(\alpha \zeta_E)d\theta^2\right]+r_+^2\left(\frac{1+\rho^2}{1-\rho^2}\right)^2d\phi^2.
\end{align}
The term in square brackets is simply the metric on $\mathbb{R}^3$ in spherical coordinates where $\zeta_E$ is the polar coordinate and is therefore non-periodic. Since imaginary time is non-periodic after a Wick rotation, this implies that the black hole is not radiating far from $\tau=0$. We will see more evidence for the freezing out of radiation in Section \ref{entanglementsec}.

\section{Action of Quotient on $D_f$ and $D_p$}

\begin{figure}
    \centering
    \includegraphics[width=10cm]{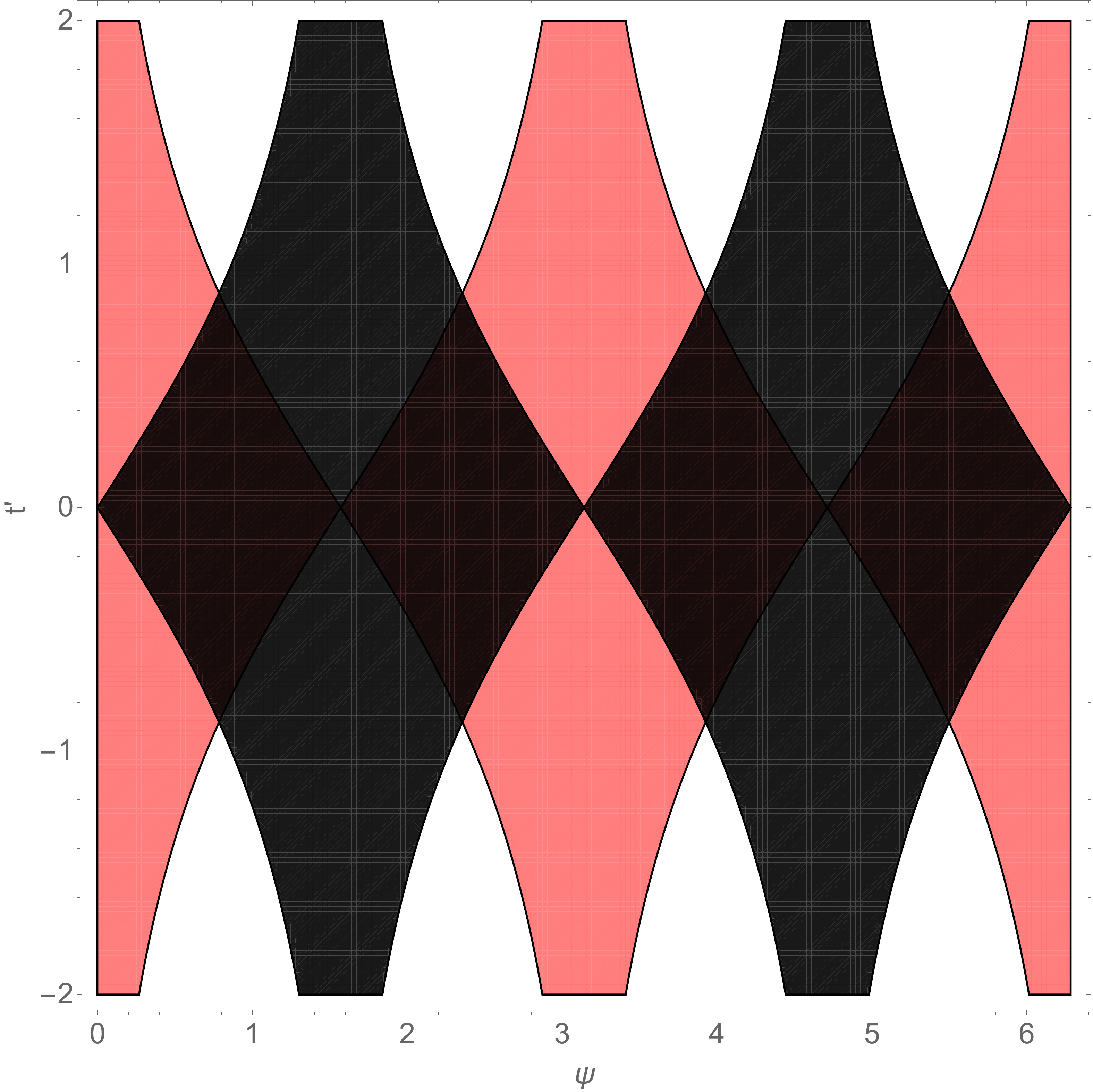}
    \caption{Regions of the boundary removed during second quotient (black) overlap the regions of the boundary in a pure state (red) at early times. However, at later times, this overlap ceases, reflecting that, at early times, the boundary CFT should be in a mixed state, but should purify at later times.}
    \label{fig:overlap}
\end{figure}

Now that we have shown how the quotient acts on the spacetime, we would next like to consider how the boundary theory behaves under this quotient. Specifically, we wish to look at what regions of the boundary are excised by the quotient. As we have seen, the quotient removes the portions of the spacetime corresponding to $y_2^2-t^2>0$. We have already found that $D_f$ and $D_p$ are the regions of the boundary $y_1^2-t^2>0$. Since $y_1\to y_2$ is the same as taking $\psi\to \psi+\pi/2$, by (\ref{fullexttrans}), it is clear that the region of the boundary removed by the quotient is given by

\begin{equation}
    \cosh^2t'\,\sin^2\psi-\sin^2t'>0
\end{equation}
in coordinates (\ref{fullext}). These removed regions overlap $D_f$ and $D_p$ at early times, as shown in Figure \ref{fig:overlap}.

The fact that we have removed regions of the boundary which can house CFTs in pure states tells us that, at early times, the CFT should, in general, be in a mixed state. However, at some later time, this overlap ceases and the states should purify.

\section{Entanglement of Boundary CFT}
\label{entanglementsec}
We have shown that the boundary CFT begins in a mixed state, but purifies after some time. We now consider what state the CFT is initially in and explicity show this purification through the use of holographic entanglement.

\begin{figure}
 \centering
 \includegraphics[width=.5\textwidth]{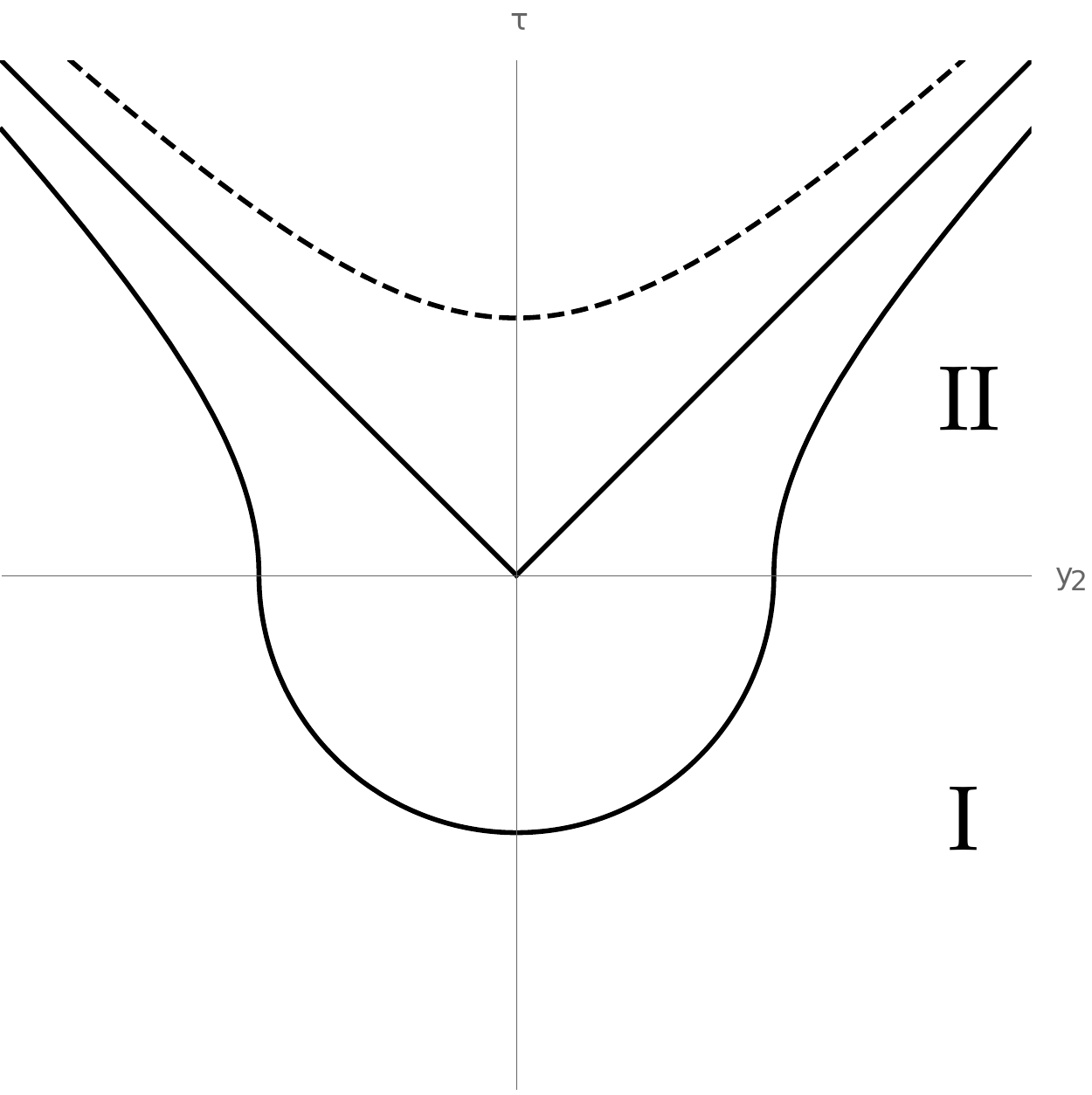}
 \caption{Hartle-Hawking initial state of the big bang spacetime. Region I is a 3-dimensional Euclidean space, where Region II is a $(3+1)$-dimensional Lorentzian spacetime. In Region I, the $\phi$ dimension is suppressed, and in Region II, the $\phi$ and $\theta$ dimensions are suppressed. Despite the spaces having different dimensionality, we can smoothly glue them along $\tau=0$ since the $\theta$ dimension collapses as $\tau\to0$.}
 \label{harthawk}
\end{figure}

Since the metric (\ref{finmet}) exhibits a symmetry under $\tau\to-\tau$, we should be able to assign a Hartle-Hawking initial condition to the spacetime \cite{hhwf}. However, we cannot obtain a Euclidean-signature metric from a single Wick rotation, $\tau\to-i\tau$. Instead, we will utilize the fact that as $\tau\to0$, the $\theta$ coordinate collapses and the metric becomes that of BTZ. We can therefore smoothly glue the 3-dimensional Euclidean metric of the BTZ spacetime, given by
\begin{align}
ds_E^2=\frac{4l^2}{\left(1-\tau_E^2-y_2^2\right)^2}&\left(d\tau_E^2+dy_2^2\right)+r_+^2\left(\frac{1+\tau_E^2+y_2^2}{1-\tau^2-y_2^2}\right)d\phi^2,
\label{emet}
\end{align}
to the metric (\ref{finmet}) along $\tau=0$ as shown in Figure \ref{harthawk}.

Since the initial condition of the spacetime is identical to that of BTZ, the state of the CFTs will also be identical very near $\tau=0$. Namely they will be in a thermofield double state, given by
\begin{equation}
\ket{\Psi}=\frac{1}{\sqrt{Z(\beta)}}\sum_n e^{-\beta E_n/2}\ket{E_n}_{D_f}\otimes \ket{E_n}_{D_p}
\end{equation}
where $D_f$ and $D_p$ refer to the two CFTs on each region of the boundary, respectively \cite{eternalbh}. We again see that at early times, the CFTs are in a mixed state, and furthermore, are entangled with each other. More importantly, this mixed state can be interpreted as thermal, a direct reflection of the fact that the spacetime is thermal at early times.

To see how the entanglement between the CFTs on $D_f$ and $D_p$ evolves, we will use the same procedure as in Section \ref{firstquosec}.

The coordinates in (\ref{finmet}) are related to Poincar\'{e} coordinates by
\begin{align}
W_{\pm}&=\frac{\pm 2 l\left|\tau\right|}{1-\tau^2+y_2^2}e^{\frac{r_+}{l}\phi\pm \alpha \theta} \; , \;
Y=\frac{2 l y_1}{1-\tau^2+y_1^2}e^{\frac{r_+}{l}\phi} \; , \;
Z=l\left(\frac{1+\tau^2-y_1^2}{1-\tau^2+y_1^2}\right)e^{\frac{r_+}{l}\phi}.
\end{align}
The boundary of the spacetime will be located at $Z=0$ or $y_1=\sqrt{1+\tau^2}$\footnote{Since the asymptotic boundaries are distinct, the region $A$ can only exist on one of these boundaries, so we choose the one corresponding to $y_1,\tau>0$.}. We can now choose to anchor the surface, $\gamma$, to the boundary at fixed time $\tau=\tau_0$, such that $\theta\in[\theta_1,\theta_2]$ and $\phi\in[\phi_1,\phi_2]$, the cutoff $Z=\epsilon\ll1$ will be given by
\begin{equation}
\epsilon_{1,2}=la\,e^{\frac{r_+}{l}\phi_{1,2}}
\label{cut}
\end{equation}
where $0<la\ll1$ and $\epsilon_{1,2}\to 0$ corresponds to $a\to 0$. Similar to above, we will center the strip such that $\theta_1=-\Theta$ and $\theta_2=\Theta$. Then, the width of the strip is given by
\begin{equation}
R^2=l\left(e^{\frac{r_+}{l}\phi_2}-e^{\frac{r_+}{l}\phi_1}\right)^2\left[1-\tau_0^2\sinh^2\left(\alpha\Theta\right)\right]
\label{width}
\end{equation}
and the length is given by
\begin{equation}
L=l\left|\tau_0\right|\left(e^{\frac{r_+}{l}\phi_2}+e^{\frac{r_+}{l}\phi_1}\right)\sinh\left(\alpha \Theta\right).
\label{length}
\end{equation}
The width, $R$, vanishes either when $\phi_1=\phi_2$ or $\tau_0\sinh(\alpha\Theta)=1$.

Using these expressions in (\ref{poinarea}), we obtain an expression for the area of the minimal area surface in coordinates (\ref{finmet})
\begin{align}
\mathrm{Area}(\gamma)=&4l^2\frac{\left|\tau_0\right|}{a}\sinh\left(\alpha\Theta\right) \, \cosh \Delta-\kappa l^2 \left(\frac{\left|\tau_0\right|\sinh\left(\alpha\Theta\right)}{\sqrt{1-\tau_0^2\sinh^2\left(\alpha\Theta\right)}}\right)\coth\Delta.
\end{align}
We can see that, as $a\to 0$ and $\tau_0\sinh\left(\alpha\Theta\right)\to1$, the area goes to zero and the region on the boundary is pure. This implies that there is a minimum time for a region of the boundary to reach a pure state
\begin{equation}
\tau_{min}=\frac{1}{\sinh\left(\alpha \Theta\right)}.
\end{equation}
When $\Theta=\pi$, the region, $A$, will be one full future asymptotic boundary. Therefore, the minimum time, $\tau_{crit}$, for the CFT on this copy of the future boundary of the spacetime to purify will be
\begin{equation}
\tau_{crit}=\frac{1}{\sinh\left(\alpha \pi\right)}.
\label{tcrit}
\end{equation}
For $\tau<\tau_{crit}$, the boundary of either future exterior is in a mixed state, which we know from the Hartle-Hawking state to be a thermofield double. For $\tau>\tau_{crit}$, the boundaries of both future exteriors are in pure states. Again, this is a reflection of the fact that, for early times, the spacetime is thermal but cools after a certain time.

We can also gain information from this about how the thermodynamic phase transition of the spacetime depends on the scaling parameter $\alpha$. Na\"{i}vely, by the form of the metric (\ref{finmet}), we may expect this phase transition to occur when the $\theta$ dimension grows from microscopic to macroscopic at times

\begin{equation}
 \tau\sim\frac{1}{\alpha}.
\end{equation}
For small values of $\alpha$, (\ref{tcrit}) appears to confirm this scaling relation. However, for $\alpha\gtrsim 1$, the boundary purifies much faster than $1/\alpha$, suggesting that the thermodynamics of the spacetime are much more sensitive to this scaling parameter than the metric would lead us to believe.

\section{Conclusions}

We have arrived at a black hole spacetime from a double quotient of global AdS$_4$ which has a time-dependent, expanding compact dimension. At early times, when this spatial dimension is small, the black hole emits thermal radiation. During this period, we also see entanglement between two of the boundary CFTs. After a critical time, the radiation stops and the entanglement is broken between the CFTs. For slow expansion, the critical time is inversely proportional to the expansion rate of the compact dimension, a result we might expect from the form of the metric. However, for rapid expansion, we have found that the purifcation time is much shorter than this na\"{i}ve inverse relation.

This work further exemplifies the power of the AdS/CFT correspondence. Using relativity or quantum mechanics alone, it is only possible to study the early and late time limits of the spacetime, but not the transition period. However, this is the epoch in which the spacetime exhibits particularly interesting characteristics. Through the application of the holographic entanglement entropy calculation, we were able to uncover specific behaviors of this transition period. We also see from the holographic calculation that the intuition we may have from just the spacetime metric alone is not necessarily correct. Moreover, this work further confirms the holographic entanglement entropy conjecture, as the holographic calculation agreed with both the early and late time limits in the relativistic and quantum calculations.

\end{document}